\documentclass{jaa}
\usepackage{natbib}

\usepackage{graphicx}
\usepackage{dcolumn}
\usepackage{bm}

\usepackage{aasmacros}

\usepackage[table,xcdraw,svgnames]{xcolor}
\usepackage[colorlinks]{hyperref}
\hypersetup{citecolor=Blue, linkcolor=Red, urlcolor=Blue}

\begin{document}\sloppy

\title{Interference patterns for simple lens models in wave-optics regime}

\author{Ashish Kumar Meena}
\affilOne{Physics Department, Ben-Gurion University of the Negev, P.O. Box 653, Be'er-Sheva 8410501, Israel}

\twocolumn[{
\maketitle
\vspace{1em}
\corres{\href{mailto:ashishmeena766@gmail.com}{ashishmeena766@gmail.com}}
\msinfo{xx yy zzzz}{xx yy zzzz}

\begin{abstract}
This work studies interference patterns created by simple lens models~(point mass, Chang-Refsdal, and binary lens) in the wave optics regime, primarily in the context of lensing of gravitational waves~(GWs) in the LIGO band at frequencies around~$100$~Hz. We study how the interference patterns behave close to the caustic curves, which mark the high magnification regions in conventional geometric optics. In addition, we also look at the formation of highly de-amplified regions in the amplification maps close to caustics and how they differ under wave and geometric optics. We see that, except for close to caustic regions, the geometric optics track the oscillations of the amplification factor in frequency very well, although the amplitude of these oscillations can differ considerably. Chang-Refsdal and binary lenses with masses~${\sim}100\:{\rm M_\odot}-200\:{\rm M_\odot}$ can introduce significant de-amplification at frequencies~${\sim}100$~Hz when the source is close to caustics, which may help us distinguish such lenses from the point mass lens.

\end{abstract}

\keywords{gravitational lensing -- gravitational waves.}
}]

\section{Introduction}
\label{sec:intro}
With the continuous detection of gravitational waves~(GWs) by the LIGO-Virgo-KAGRA (LVK) network~\citep[e.g.,][]{2023PhRvX..13d1039A, 2024PhRvD.109b2001A} and the fact that gravitational lensing by stellar mass objects can introduce chromatic~(i.e., frequency-dependent) effects in the corresponding frequency range, it has become crucial to understand gravitational lensing effects in the wave optics regime. Generally, wave effects become important if the lens mass is~$M\lesssim10^5(f/{\rm Hz})^{-1}$, where~$f$ is the frequency of the incoming signal~\citep[e.g.,][]{2003ApJ...595.1039T, 2022JCAP...07..022B}. More specifically, for GW signals in the LVK frequency range, the above mass primarily contains microlenses~(e.g., stars and stellar mass compact objects) and milli-lenses~(e.g., star clusters and sub-halos).

Under geometric optics approximation, a background GW source is multiply imaged, and for a given source position, the contribution to the amplification factor~\citep[also known as transmission factor; e.g.,][]{1999prop.book.....B} only comes from stationary points of the Fermat potential~\citep[e.g.,][]{1986ApJ...310..568B}, which essentially denote positions of multiple images in the image plane. In generic lensing scenarios under geometric optics, the time delays between these multiple images are large~(ranging from days to years) compared to the length of the GW signal~(${\lesssim}1$~second) in the LVK frequency band, and we expect to observe one image at a time, (de-)amplified (and phase-shifted) by a constant factor. Once we go into the wave optics regime, we do not observe these multiple lensed images. Although, as we will see below, images formed in geometric optics~(i.e., virtual images) are still relevant in understanding various features in the wave optics regime. In wave optics, the incoming wavefront is treated as a constant phase surface, and the lens (plane) can be thought of as a transparent phase screen introducing an extra phase (which depends on the frequency and lens parameters) for every partial wave of the incoming wavefront~\citep[e.g.,][]{1992grle.book.....S, 1999prop.book.....B, 2022JCAP...07..022B}, and the observer detects the resulting interference pattern, including contributions also from non-stationary points. Due to this extra contribution in wave optics, the resulting interference pattern is also sensitive to the lens model properties and can help break various lensing degeneracies~\citep{2023PhRvD.108d3527T, 2024PhRvD.110j3024M, 2025MNRAS.536.2212P}.

Most of the previous works have focused on studying the wave effects assuming a fixed source (or observer) position and as a function of frequency for an isolated point mass (i.e., Schwarzschild) lens~\citep[e.g.,][]{2003ApJ...595.1039T, 2022JCAP...07..022B}, from a point mass lens in the presence of external effects~\citep[e.g.,][]{2020MNRAS.492.1127M, 2021MNRAS.508.4869M, 2023MNRAS.526.2230Y}, and from an ensemble of point mass lenses equivalent to microlens population in galaxy and galaxy cluster lenses~\citep[e.g.,][]{2019A&A...627A.130D, 2021MNRAS.508.4869M, 2022MNRAS.517..872M, 2023SCPMA..6639511S, 2024MNRAS.532.3568M}. This can be understood from the fact that, realistically, an observed lensed GW signal would span a range of frequencies and have one source position. In addition, at present, to our knowledge, we do not have efficient methods to simulate a large number of amplification factors for different source positions\footnote{Except for a point mass lens for which we have an analytical formula for the amplification factor.}. This can limit us from visualising the effect of the variation of the source position on the amplification factor at a given frequency, especially close to the caustics, where extra virtual images (under the geometric optics approximation) appear or disappear.

In our current work, we simulate interference patterns/maps in the source plane for simple lens models (point mass, Chang-Refsdal, and binary lens), primarily focusing on GW signals in the LVK frequency range. Strictly speaking, we simulate maps of the absolute value of the amplification factor. However, an interference pattern is essentially a map of~$\rm |amplification|^2$ and we use the above terminologies interchangeably. For the point mass lens case, thanks to the circular symmetry, the source position can be given by its polar coordinate, and an amplification map can be constructed on the source position and dimensionless frequency plane, which can be re-scaled for a given lens mass~\citep[see Fig. 4 in][]{2022JCAP...07..022B}. However, the same cannot be done once we break the circular symmetry, as we have three-dimensional space with two parameters describing the source position and one for (dimensionless) frequency. Hence, for simplicity, we choose three frequency values~$f=50,100,500$~Hz to show interference maps as they cover the range of frequencies where nearly all signals detected by the LVK network are expected to have non-zero powers. We also discuss how our results can be used for other frequencies and lens masses. With these simulated maps, we study how the interference map varies as the number of virtual images changes close to caustics. It is possible that for certain lens system configurations~(or source positions assuming everything else is fixed), different parts of the incoming wavefront interfere destructively and lead to highly de-amplified regions. Under geometric optics, finding such configurations is easy, and it would be interesting to see how closely geometric optics follows the full wave optics solution. Hence, we also look at the formation of such regions in the source plane under wave optics and at what lens mass range and source positions they occur at frequencies~${\sim}100$~Hz so that we may detect them in LVK signals.

The current work is organized as follows. Section~\ref{sec:gl_basic} briefly reviews the relevant lensing basics in wave optics. Section~\ref{sec:Ff_calc} describes the method we employ in our current work to evaluate the amplification factor. Section~\ref{sec:pml},~\ref{sec:crl},~\ref{sec:bml} study interference patterns for point, Chang-Refsdal, and binary lens models, respectively. We summarise this work in Section~\ref{sec:summary}.

\begin{figure*}[!ht]
    \centering
    \includegraphics[width=17.5cm,height=5.8cm]{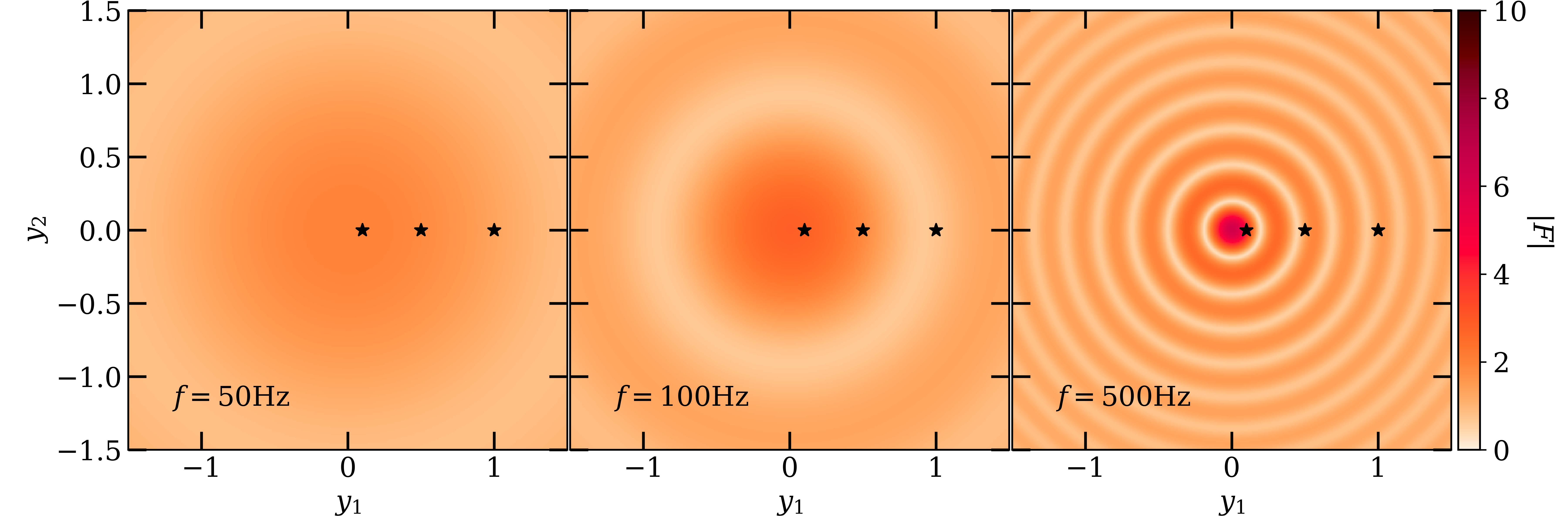}
    \includegraphics[width=17.5cm,height=5cm]{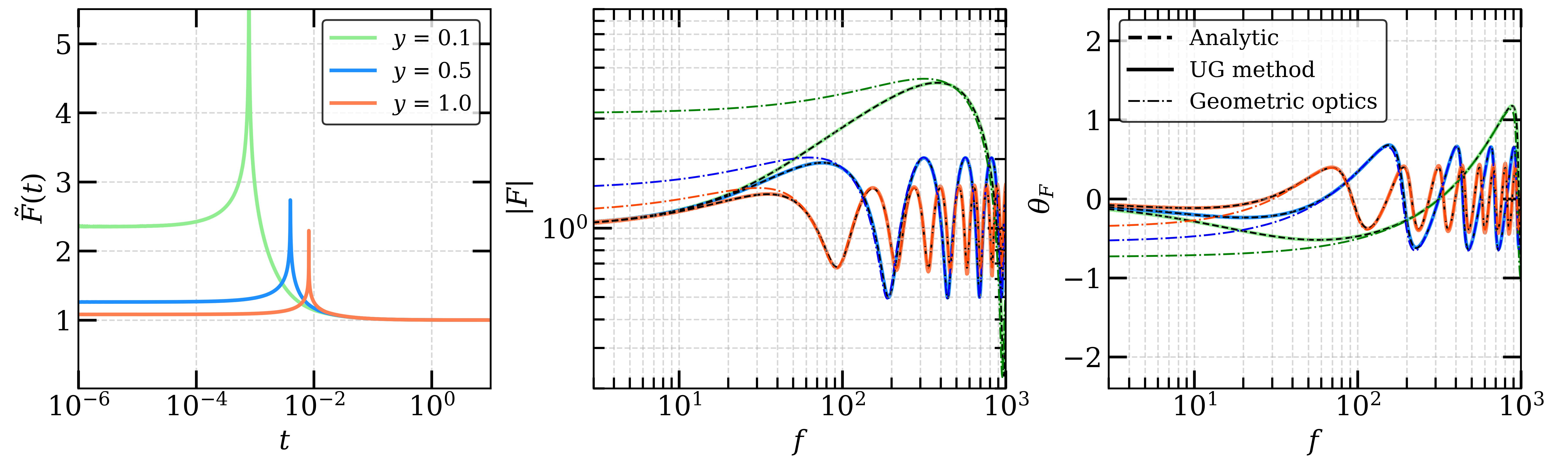}
    \caption{Amplification factor, $F(f)$, for point mass lens. 
    \textit{Top-row}: Source plane maps of the absolute value of amplification factor~($|F|$) at three different frequencies. 
    \textit{Bottom-left}: $\Tilde{F}(t)$ curves normalized by~$2\pi$ for three different values of source positions marked by the black solid stars in the top row.
    \textit{Bottom-middle}: Absolute value of amplification factor~($|F|$). The black dashed curves represent $|F|$ calculated using the analytical formula in Equation~\eqref{eq:pm_analytic}, and coloured solid curves are calculated using the UG method. The dashed-dotted curves show the~$|F|$ under geometric optics approximation.
    \textit{Bottom-right}: Phase value of amplification factor~($\theta_F$). Like the middle panel, the black dashed curves are estimated using the analytic formula. Coloured solid curves are calculated using the UG method, and dashed-dotted curves are the results under the geometric optics approximation.}
    \label{fig:pm}
\end{figure*}

\section{Lensing basics in wave optics regime}
\label{sec:gl_basic}
This section briefly revisits the relevant basics of gravitational lensing in wave optics regimes. For a more detailed description, we refer readers to the following works:~\citet{1992grle.book.....S, 1999PThPS.133..137N, 2003ApJ...595.1039T}.

The amplification factor, which is defined as the ratio of GW amplitude in the presence and absence of the lens, is given as~\citep[e.g.,][]{1999PThPS.133..137N, 2003ApJ...595.1039T},
\begin{equation}
    F(f,\pmb{y}) = \frac{1+z_d}{\rm c} \frac{D_d D_s}{D_{ds}} \theta_0^2 \frac{f}{i} \int d^2\pmb{x} \exp[2\pi i f t_d(\pmb{x}, \pmb{y})],
    \label{eq:Ff_gen}
\end{equation}
where~$z_d$ is the lens redshift. $\pmb{x}\equiv\pmb{\theta}/\theta_0$ and~$\pmb{y}\equiv\pmb{\beta}/\theta_0$ are the dimensionless position vectors in the image and source plane, respectively, with~$\theta_0$ being the normalization angular scale. $D_d$, $D_s$, and $D_{ds}$ are angular diameter distances from observe to the lens, observer to source, and lens to source, respectively. $t_d(\pmb{x}, \pmb{y})$ represents the well-known arrival time delay surface given as,
\begin{equation}
    t_d(\pmb{x}, \pmb{y}) = \frac{1+z_d}{c} \frac{D_d D_s}{D_{ds}} \theta_0^2 
                            \left[ \frac{(\pmb{x} - \pmb{y})^2}{2} - \psi(\pmb{x}) + \phi_m(\pmb{y}) \right],
    \label{eq:td_gen}
\end{equation}
where~$\psi(\pmb{x})$ is the lensing potential and~$\phi_m(\pmb{y})$ is a constant independent of lens properties, which is usually set to a value such that the arrival time delay for global minima is zero. As we can see from the above equation, the amplification factor,~$F(f)$, is a frequency-dependent complex-valued function. For GWs, the change in amplitude of the signal due to lensing is given by the corresponding absolute value~$(|F|)$ and lensing-induced phase-shift is given by~$\theta_F\equiv -i \ln[F/|F|]$. In geometric optics limit~$(f t_d \gg 1)$, the integral in Equation~\eqref{eq:Ff_gen} becomes highly oscillatory, and only its stationary points contribute to the amplification factor such that
\begin{equation}
    F(f, \pmb{y})|_{\rm geo} = \sum_j \sqrt{|\mu_j|} 
    \exp \left( 2\pi i f t_d(\pmb{x}_j, \pmb{y}) - i \pi n_j \right),
    \label{eq:Ff_geo}
\end{equation}
where~$\mu_j$ is the geometric optics magnification of the $j$-th image and~$n_j$ is the Morse index with values~0, 1/2, and~1 for minima, saddle-point, and maxima images, respectively~\citep[e.g.,][]{1992grle.book.....S, 1999PThPS.133..137N}. The extra phase shift arises from complex Gaussian integral (close to stationary points) and can be used to identify lensed GW signals~\citep[e.g.,][]{2021PhRvD.103f4047E, 2021PhRvD.103j4055W, 2023PhRvD.108d3036V}.

Often, at least for simple lens models, it is useful to write the time delay and frequency in dimensionless quantities, which can be defined as~\citep{2021MNRAS.508.4869M},
\begin{equation}
    \begin{split}
        T_s &= \frac{1+z_d}{c} \frac{D_d D_s}{D_{ds}} \theta_0^2, \\
        \nu &= T_s f, \qquad \tau_d(\pmb{x}, \pmb{y}) = \frac{t_d(\pmb{x}, \pmb{y})}{T_s},
    \end{split}
\end{equation}
such that the amplification factor takes the form,
\begin{equation}
    \begin{split}
        F(\nu,\pmb{y}) &= \frac{\nu}{i} \int d^2\pmb{x} \exp[2\pi i \nu \tau_d(\pmb{x}, \pmb{y})], \\
        F(\nu, \pmb{y})|_{\rm geo} &= \sum_j \sqrt{|\mu_j|} \exp \left( 2\pi i \nu \tau_d(\pmb{x}_j, \pmb{y}) - i \pi n_j \right).
    \end{split}
    \label{eq:Ff_nu}
\end{equation}
As we will see below, such a dimensionless form allows us to translate results for one particular lens mass~($M_z$) or frequency~($f$) to another.

\section{Calculating $F(\lowercase{f})$}
\label{sec:Ff_calc}
Except for very simple lens models, one cannot obtain an analytic solution for Equation~\eqref{eq:Ff_gen} or~\eqref{eq:Ff_nu}. Hence, for a general lens, we need to rely on numerical integration or some other methods~\citep[e.g.,][]{1995ApJ...442...67U, 2019A&A...627A.130D, 2025PhRvD.111j3539V}. In our current work, we primarily use the method described in~\citet[][`UG method' hereafter]{1995ApJ...442...67U}, which relies on the Fourier transformation and contour integration. First, assuming a constant source position, we take the Fourier transformation of~$i F(\nu)/\nu$ in Equation~\eqref{eq:Ff_nu} defined as,
\begin{equation}
    \Tilde{F}(\tau') \equiv \int d\nu \: \exp(-2\pi i \nu\tau') \frac{i F(\nu)}{\nu}.
    \label{eq:Ft}
\end{equation}
After substitution from Equation~\eqref{eq:Ff_nu}, the above takes the form,
\begin{equation}
    \Tilde{F}(\tau') = \int d^2\pmb{x} \: \delta[\tau_d(\pmb{x};\pmb{y}) - \tau'].
\end{equation}
This implies that contribution to~$F(\tau')$ only comes from constant time delay contours corresponding to~$\tau'$. Taking the inverse Fourier transformation and substituting~$\nu = T_s f$, we get,
\begin{equation}
    F(f) = \frac{f}{i} \int dt_d \exp(2\pi i f t_d) \: \Tilde{F}(t_d),
    \label{eq:Ff_Ft}
\end{equation}
where~$t_d$ represents the time-delay value relative to an arbitrary reference time, which we set to be the first image to arrive. With the above, the problem transforms into determining the~$\Tilde{F}(t)$ and performing an inverse Fourier transformation.

To determine~$\Tilde{F}(t)$, we use contour integration. As discussed in~\cite{2021MNRAS.508.4869M}, the area between contours of~$\tau'$ and~$\tau' + d\tau'$ is equal to~$\oint ds dl$, where~$ds$ is the infinitesimal length along the contour and~$dl=d\tau'/|\nabla_{\pmb{x}} \tau_d(\pmb{x}, \pmb{y})|$ is the distance in the orthogonal direction between the two contours. Since there can be more than one contour with a value of~$\tau'$
\begin{equation}
    \Tilde{F}(\tau') = \sum_k  \oint_{C_k}{ \frac{ds}{|\nabla_{\pmb x} \tau_d(\pmb{x}, \pmb{y})|} },
\end{equation}
where the sum is over all contours,~$C_k$, with~$\tau_d(\pmb{x}, \pmb{y})=\tau'$. From the above equation, we can also see that~$\Tilde{F}(\tau)$ is a smooth function except at critical points where images appear, i.e., where~$|\nabla_{\pmb{x}} \tau_d(\pmb{x}, \pmb{y})|=0$. We refer the reader to~\cite{1995ApJ...442...67U} for more details on how to handle such points.

\begin{figure*}[!ht]
    \centering
    \includegraphics[width=17.5cm,height=15.8cm]{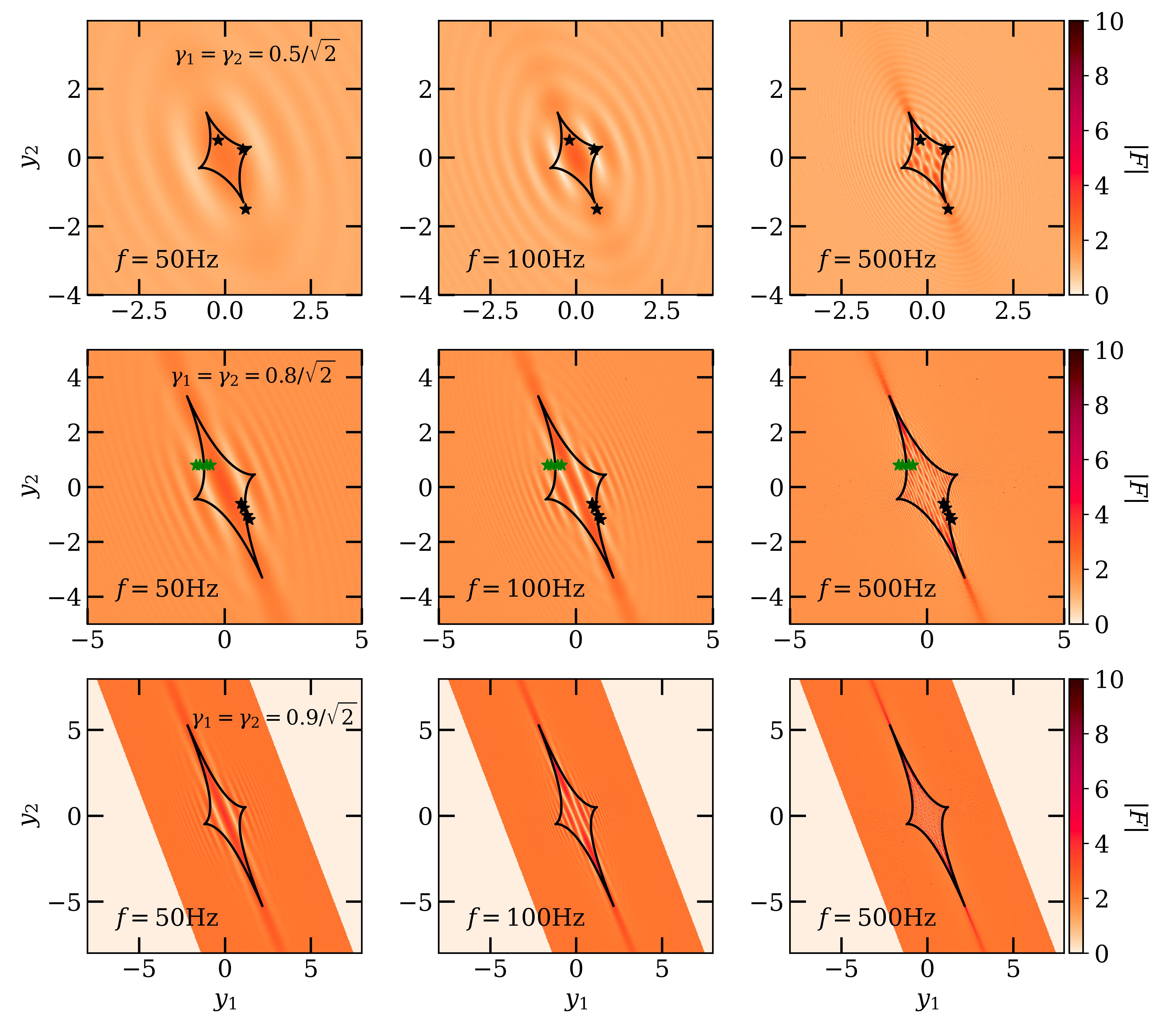}
    \caption{Amplification factor, $|F|$, maps for Chang-Refsdal lens with positive values of external shear~(i.e., $\gamma > 0$). The three rows represent the source plane maps corresponding to~$\gamma = 0.5/\sqrt{2}, 0.8/\sqrt{2}, 0.9/\sqrt{2}$~(i.e.,~$\mu_{\rm m} = 1.33, 2.78, 5.26$), respectively. In each row, left, middle and right panels correspond to~$f=50, 100, 500$~Hz. In each panel, black solid curves represent the caustic. The black and green stars in the top and middle rows represent the source position for which amplification factor curves are shown in Figure~\ref{fig:cr_Ff_curves}.}
    \label{fig:cr_interf_comb}
\end{figure*}

\begin{figure*}[!ht]
    \centering
    \includegraphics[width=17.5cm,height=5cm]{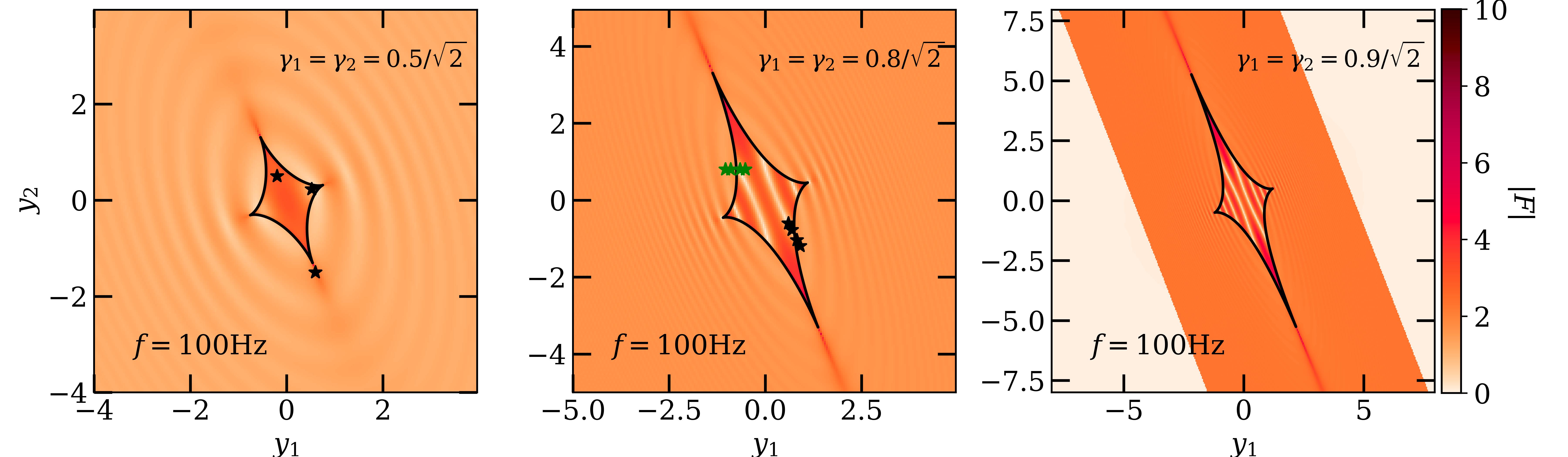}
    \caption{Amplification factor,~$|F|$, maps under geometric optics approximation for the Chang-Refsdal lens with~$M_z=200~{\rm M_\odot}$ at~$f=100$~Hz. The left, middle, and right panels correspond to three different~$\gamma$ values, which are shown in the top-right part of each panel. The corresponding interference maps of~$|F|$ in the wave optics are shown in the middle column of Figure~\ref{fig:cr_interf_comb}. The amplification factor curves for black and green stars in the left and middle panels are shown in Figure~\ref{fig:cr_Ff_curves}.}
    \label{fig:cr_interf_geo}
\end{figure*}

\begin{figure*}[!ht]
    \centering
    \includegraphics[width=17.5cm,height=5cm]{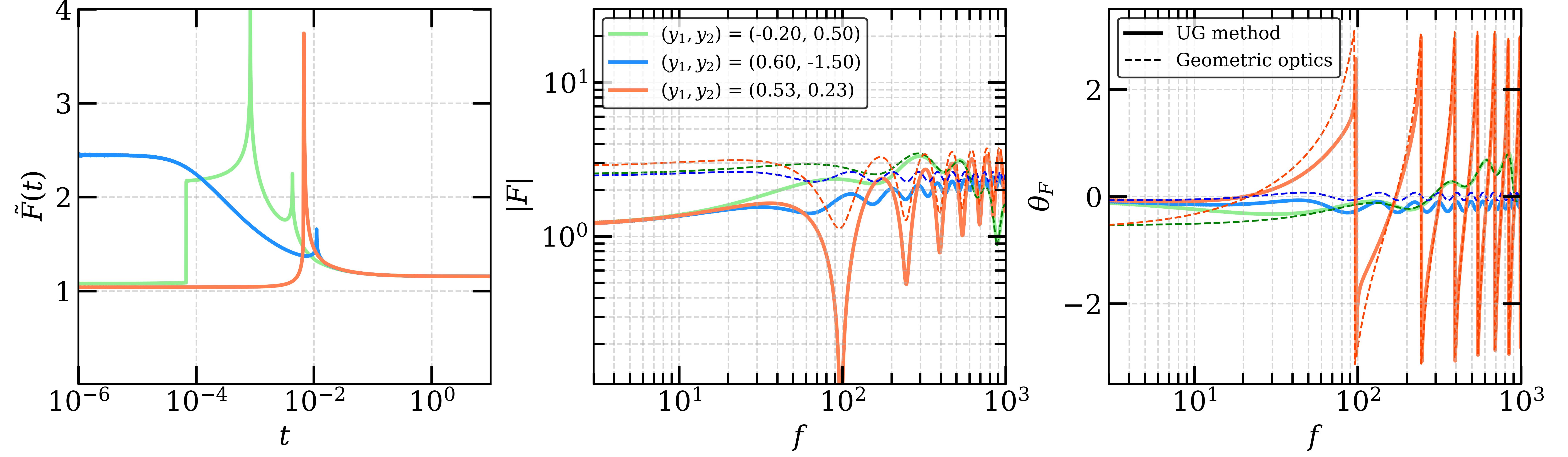}
    \includegraphics[width=17.5cm,height=5cm]{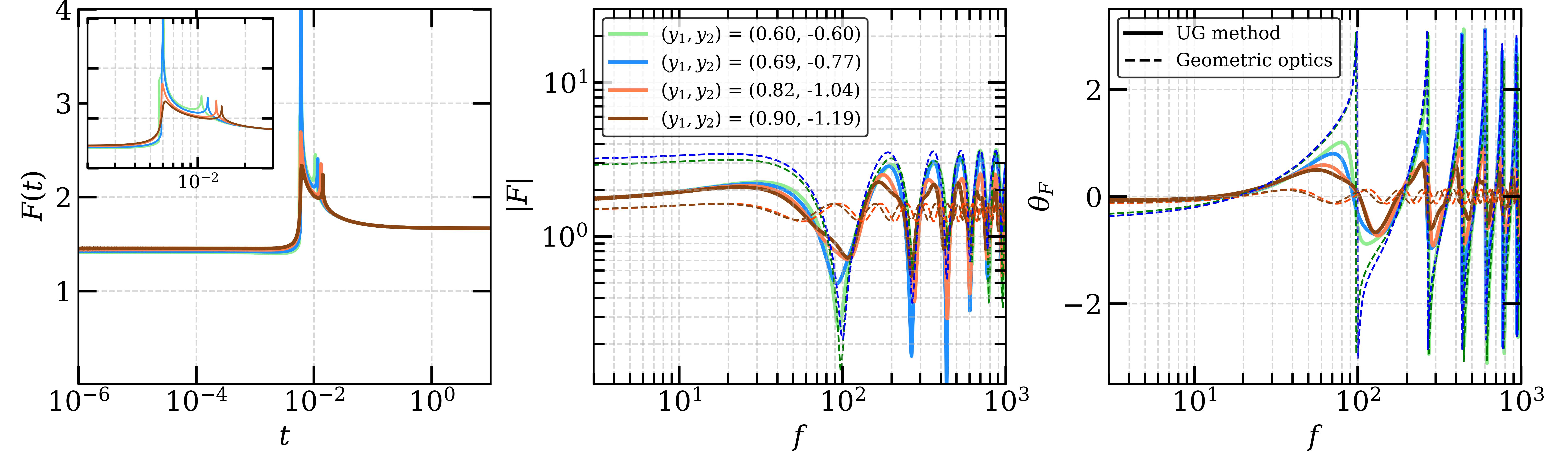}
    \includegraphics[width=17.5cm,height=5cm]{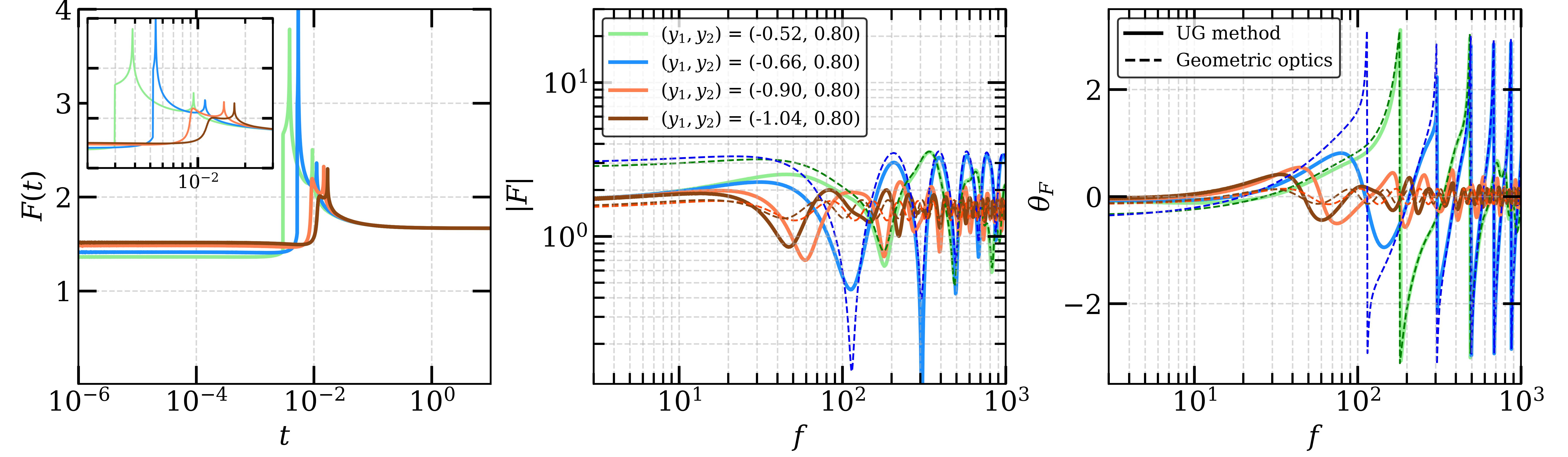}
    \caption{Example amplification factor, $F(f)$, curves for the Chang-Refsdal lens with~$\gamma = 0.5/\sqrt{2}$ in the top panel and~$\gamma = 0.8/\sqrt{2}$ in the middle and bottom panels, respectively. The source positions in the top panel are shown in the top panels of Figure~\ref{fig:cr_interf_comb} and are marked by solid black stars. Source positions for the middle and bottom panels are shown by black and green stars in the middle panels of Figure~\ref{fig:cr_interf_comb}. The left, middle, and right panels show the~$\Tilde{F}(t)$,~$|F|$, and~$\theta_F$ curves, respectively. The solid curves are obtained using the UG method and the corresponding geometric optics approximation results are shown by dashed curves.}
    \label{fig:cr_Ff_curves}
\end{figure*}

\section{Point mass lens}
\label{sec:pml}
For a point lens with mass~$M$, the lensing potential~(in dimensionless form) is given as,
\begin{equation}
    \psi(x) = \ln|x|.
\end{equation}
Here, we have assumed that the normalization angular scale~($\theta_0$) is equal to the corresponding Einstein angle~($\theta_E$), which is given as,
\begin{equation}
    \theta_E = \sqrt{\frac{4{\rm G}M}{{\rm c^2}} \frac{D_{ds}}{D_d D_s}},
\end{equation}
where various symbols have their usual meaning. The corresponding time delay function, up to an additive constant, is given as,
\begin{equation}
    t_d(x, y) = \frac{(1+z_d)}{\rm c} \frac{4{\rm G}M}{{\rm c^2}} \left[ \frac{(x - y)^2}{2} - \ln|x| \right].
\end{equation}
For the point mass lens, the amplification factor, Equation~\eqref{eq:Ff_gen}, has an analytical solution, which is given as,
\begin{equation}
    \begin{split}
    F(f,y) = \exp \left\{ \frac{\pi \omega}{4} + 
    \frac{i\omega}{2} \left[ \ln \left(\frac{\omega}{2}\right) - 2 \phi_{\rm m}(y) \right] \right\} \\
    \Gamma\left( 1-\frac{i\omega}{2} \right) {}_1F_1 \left( \frac{i\omega}{2}, 1; \frac{i\omega y^2}{2} \right),         
    \end{split}
    \label{eq:pm_analytic}
\end{equation}
where~$\omega\equiv2\pi\nu$. $\phi_{\rm m}(y)=(x_{\rm m}-y)^2/2-\ln(x_{\rm m})$ with $x_{\rm m} = (y+\sqrt{y^2+4})/2$ represents the time delay corresponding to the global minima image in geometric optics such that it arrives at~$t_d=0$. ${}_1F_1(a, b, z)$ is the confluent hypergeometric function. In geometric optics limit~($\nu \tau_d \gg 1$), the amplification factor can be written as,
\begin{equation}
    F(f,y)|_{\rm geo} = \sqrt{|\mu_+|} - i \sqrt{|\mu_-|} \exp[2\pi i \nu \tau_d(x, y)],
    \label{eq:pm_geo}
\end{equation}
where~$\mu_+$ and~$\mu_-$ are the usual magnification factors of minimum and saddle-point images, respectively.

Lensing by a point mass lens can be described by two parameters: redshifted lens mass~($M_z$) and source position~($y$). Due to the axial symmetry in the case of a point mass lens, we can expect the interference pattern to show rings of constructive and destructive interference with the distance between consecutive maxima and minima depending on the lens mass and incoming signal frequency~\citep{2022JCAP...07..022B}. The amplification factors for a point mass lens with~$M_{z}=200~{\rm M_\odot}$ are shown in Figure~\ref{fig:pm}. The top row represents the source plane map of the amplification factor~($|F|$) for three different frequencies,~$f=50, 100, 500$~Hz, and the bottom row represents the~$F(t)$, $|F|$, and $\theta_F$ for three different randomly chosen source positions. These amplification maps can also represent other lens masses or frequencies for the same source position such that they lead to the same value of dimensionless frequency~($\nu=4{\rm G}M_z/{\rm c^3}$).

In the case of an isolated point mass lens, we always have the formation of two images in geometric optics, with the saddle-point image becoming less and less amplified (approaching zero), and the minima image amplification factor approaches one as the source moves away from the optical axis~(i.e., origin). At the same time, the time delay between the two images also increases. Using the geometric optics approximation from Equation~\eqref{eq:pm_geo}, we can see that the increase in time delay implies that oscillations in amplification factor would become more rapid at a given frequency, and we move towards geometric optics at relatively lower frequencies. The larger fractional difference in the amplification of two images decreases the amplitude of the oscillations in the amplification factor. Both of the above can be seen from the dashed-dotted curves in the bottom row of Figure~\ref{fig:pm}. In addition, we can also see that the amplification factor under geometric optics approximation and wave optics mainly differs at frequencies lower than the first peak, implying that above the frequency of the first peak, the contribution to the amplification factor is well localized around the stationary points. For the perfect alignment~(i.e., $y=0$) case, we have the formation of a ring with zero time delay, implying that we reach geometric optics at~$f=\infty$ and wave effects are important for every observable frequency.

\begin{figure*}[!ht]
    \centering
    \includegraphics[width=17.5cm,height=20.8cm]{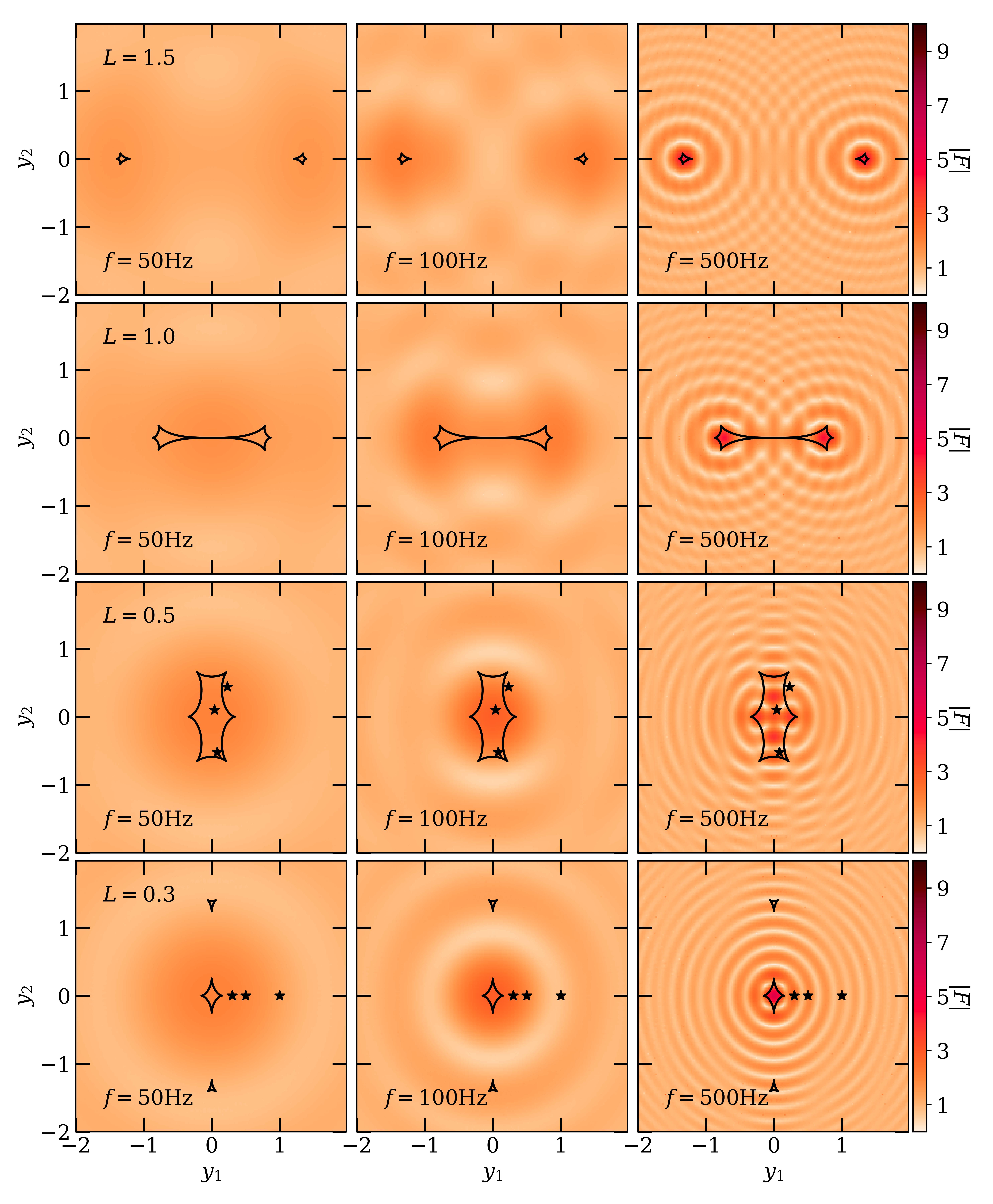}
    \caption{Amplification factor, $|F|$, maps for binary lens with~$(q_1, q_2) = (0.5, 0.5)$ and~$M_T=200~{\rm M_\odot}$. In each row, left, middle, and right panels are corresponding to~$f=50, 100, 500$~Hz. From top-to-bottom, we decrease the distance (2$L$) between the two-point masses. In each panel, black solid curves represent the caustic structure. The black solid stars are source positions for which we show the amplification factor in Figure~\ref{fig:bml_Ff}.}
    \label{fig:bml_interf_comb1}
\end{figure*}

\begin{figure*}[!ht]
    \centering
    \includegraphics[width=17.5cm,height=5cm]{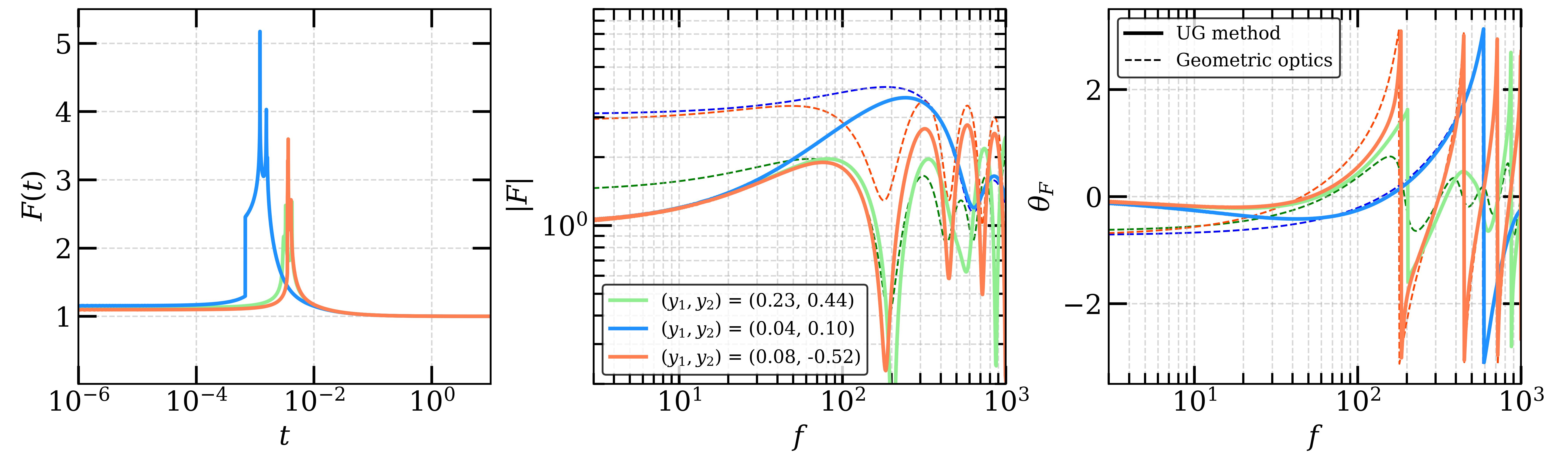}
    \includegraphics[width=17.5cm,height=5cm]{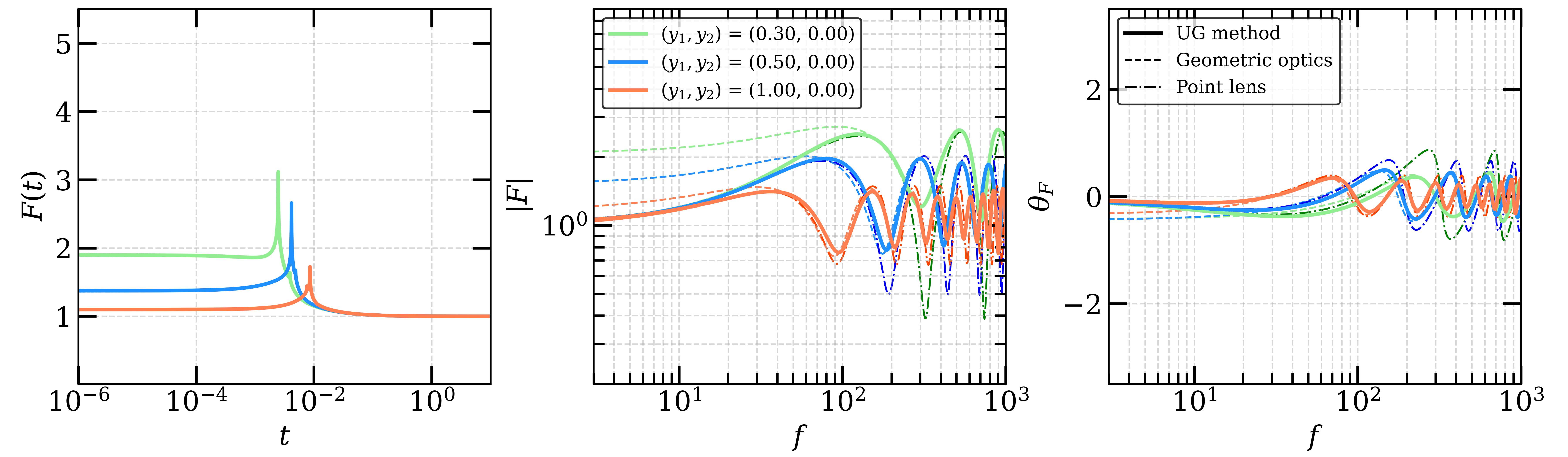}
    \caption{Example amplification factor,~$F(f)$, for binary lens case. Source positions are shown by solid stars in the third and fourth rows of Figure~\ref{fig:bml_interf_comb1}. The left, middle, and right panels show the~$F(t)$,~$|F|$, and~$\theta_F$ curves, respectively. The solid curves are obtained with the UG method and dashed curves represent the amplification factors obtained under geometric optics. In the bottom panel, for comparison, the dashed-dotted curves represent the amplification factor curves for a point mass lens with~$M_z=200~{\rm M_\odot}$ and~$y=|\pmb{y}|$.}
    \label{fig:bml_Ff}
\end{figure*}

\section{Chang-Refsdal lens}
\label{sec:crl}
After the point mass lens, the next simplest lens model is a point mass in the presence of external shear caused by some external mass distribution, also known as the Chang-Refsdal lens~\citep{1979Natur.282..561C, 1984A&A...132..168C}. The lensing potential (in dimensionless form) for such a lens model is given as,
\begin{equation}
    \psi(\pmb{x}) = \ln|\pmb{x}| 
    + \frac{\gamma_1'}{2} \left(x_1^2 - x_2^2\right) 
    + \gamma_2' x_1 x_2,
    \label{eq:cr_pot}
\end{equation}
where~$(\gamma_1', \gamma_2')$ denote the components of the external shear. Here, we have assumed that the point mass lens is at the origin. With this, the arrival time delay surface takes the form,
\begin{equation}
    \begin{split}
    t_d(\pmb{x}, \pmb{y}) = \frac{(1+z_d)}{\rm c}\frac{4{\rm G}M}{{\rm c^2}} 
    \biggr[ \frac{(\pmb{x} - \pmb{y})^2}{2} - \ln|\pmb{x}| \\ 
     - \frac{\gamma_1'}{2} \left(x_1^2 - x_2^2\right) - \gamma_2' x_1 x_2 \biggr],
    \end{split}
    \label{eq:cr_td}
\end{equation}
where~$M$ is the mass of the point lens, and other quantities have their usual meaning. In principle, we can always choose a coordinate system such that in this new coordinate system~$\gamma_1' \equiv \gamma = \sqrt{\gamma_1'^2+\gamma_2'^2}$ and~$\gamma_2' = 0$. Hence, the Chang-Refsdal lens is described by a total of four free parameters,~$(M_z, \gamma, y1, y2)$. However, in the current work, we choose~$\gamma_1' = \gamma_2' = \gamma/2$ so that caustics are somewhat diagonal in the plots for better visualization. In the absence of a point mass lens, the lensing (macro-)magnification is given as~$\mu_{\rm m} = 1/(1-\gamma^2)$ and based on the value of~$\gamma$ the Chang-Refsdal lens represents a point mass lens sitting close to a positive parity~($\gamma<1$) or negative parity~($\gamma>1$) image formed due to the external mass distribution. Note that the positive parity case~($\gamma<1$) only corresponds to the minimum image. For the maximum image, one also needs to add a constant convergence~($\kappa'$) such that both eigenvalues of the lensing Jacobian are negative. For~$\kappa'>1$, sometimes also referred to as the over-focussing case, it is possible to have a region in the source plane where the number of lensed images is zero~\citep[e.g.,][]{1984A&A...132..168C, 2006MNRAS.369..317A} and it would be interesting to see how the amplification maps look compared to other cases. In this work, for simplicity, we limit ourselves to~$\gamma<1$ (i.e.,~minimum image) and other cases are left for the future.

The amplification~($|F|$) maps for three different values of~$\gamma$ are shown in Figure~\ref{fig:cr_interf_comb} for~$f=50, 100, 500$~Hz. We can see that, due to the breaking of circular symmetry, the oscillations in the maps are no longer circular and align with the shear direction~(or major axis of diamond caustic). For comparison, the corresponding amplification maps for~$f=100$~Hz~(i.e., middle column in Figure~\ref{fig:cr_interf_comb}) under geometric optics are shown in Figure~\ref{fig:cr_interf_geo}. Comparing Figure~\ref{fig:cr_interf_comb} and~\ref{fig:cr_interf_geo}, we can see that inside the diamond caustic as well as far from it in outer regions, the positions of de-amplified regions are well-traced even under the geometric optics approximation. As we cross the fold caustic (from inside to outside), two of the extrema points of the arrival time delay surface merge and disappear, which explains the sudden change in the~$|F|$ maps under geometric optics approximation in Figure~\ref{fig:cr_interf_geo}. On the other hand, considering the full wave-optics solution, we can see that there is still a non-zero contribution from regions close to the points where the two images merged, as the distortion of the arrival time delay surface is a continuous function of the source position. In addition, we can also appreciate the effect of macro-magnification,~$\mu_{\rm m}$. An increase in~$\mu_{\rm m}$ effectively brings the oscillations in~$F(f)$ at lower frequencies.

Example amplification curves for source positions marked by black solid stars in the top row, black solid stars in the middle row, and green solid stars in the middle row of Figure~\ref{fig:cr_interf_comb} are shown in the top, middle, and bottom rows of Figure~\ref{fig:cr_Ff_curves}, respectively. For the source outside the diamond caustic and close to the cusp point in the top row of Figure~\ref{fig:cr_interf_comb}, we have the formation of two virtual images with the minima being highly amplified. And as we saw in the point mass lens case, interference of two images with very different amplification factors, although it leads to frequency-dependent oscillations, the amplitude of these oscillations is small, as seen in the blue curve in the top row of Figure~\ref{fig:cr_Ff_curves}. We also notice that solutions under geometric optics differ significantly compared to the full wave optics solution. This difference between geometric and wave optics solutions is even more prominent for the orange curve representing a source inside the diamond caustic, which has three images with high magnification factors in geometric optics. That said, in the top panel of Figure~\ref{fig:cr_Ff_curves}, we see that the position of crests and troughs along the frequency axis matches in wave and geometric optics.

In the middle row of Figure~\ref{fig:cr_Ff_curves}, we show amplification factors for source positions such that they follow a de-amplified region around the diamond caustic along the major axis, whereas the bottom row represents sources that go perpendicular to the de-amplified region. The merger and disappearance of images can be seen in the~$\Tilde{F}(t)$ curves and the corresponding inset plots. Here, for a source inside the diamond caustic, we again see that the position of crests and troughs along the frequency axis matches in wave and geometric optics. However, once we cross the caustic, the geometric optics solution loses contribution from two images and differs significantly from the full wave optics solution.

\section{Binary lens}
\label{sec:bml}
For a binary lens with masses~$M_1$ and~$M_2$, the lensing potential can be written as,
\begin{equation}
    \psi(\pmb{x}) = q_1 \ln|\pmb{x}-\pmb{L}| + q_2 \ln|\pmb{x}+\pmb{L}|,
    \label{eq:bml_pot}
\end{equation}
where~$q_i=M_i/M_T$ with~$M_T=M_1+M_2$ represents the mass ratios of primary and secondary components and~$2|\pmb{L}|$ is the distance between the two lens components. Here, we have assumed that both point mass lenses are at equal distances from the origin. The corresponding arrival time delay surface is given as,
\begin{equation}
    \begin{split}
    t_d(\pmb{x}, \pmb{y}) = \frac{(1+z_d)}{\rm c}\frac{4{\rm G}M_T}{{\rm c^2}} 
    \biggl[ \frac{(\pmb{x} - \pmb{y})^2}{2} - \: q_1\ln|\pmb{x}-\pmb{L}|  \\ 
     - \: q_2\ln|\pmb{x}+\pmb{L}| \biggr],      
    \end{split}
    \label{eq:bml_td}
\end{equation}
where various quantities have their usual meaning. Without loss of generality, we can always choose a coordinate system such that both lenses lie on the x-axis, i.e., $\pmb{L} = (L,0)$. With that, we can see that a binary lens can be described by a total of five parameters,~$(M_{1z}, M_{2z}, L, y_1, y_2)$. Note that the Chang-Refsdal lens is also a limiting case of the binary lens~\citep{1986A&A...164..237S, 1999A&A...349..108D}.

\begin{figure*}[!ht]
    \centering
    \includegraphics[width=17.5cm,height=20.8cm]{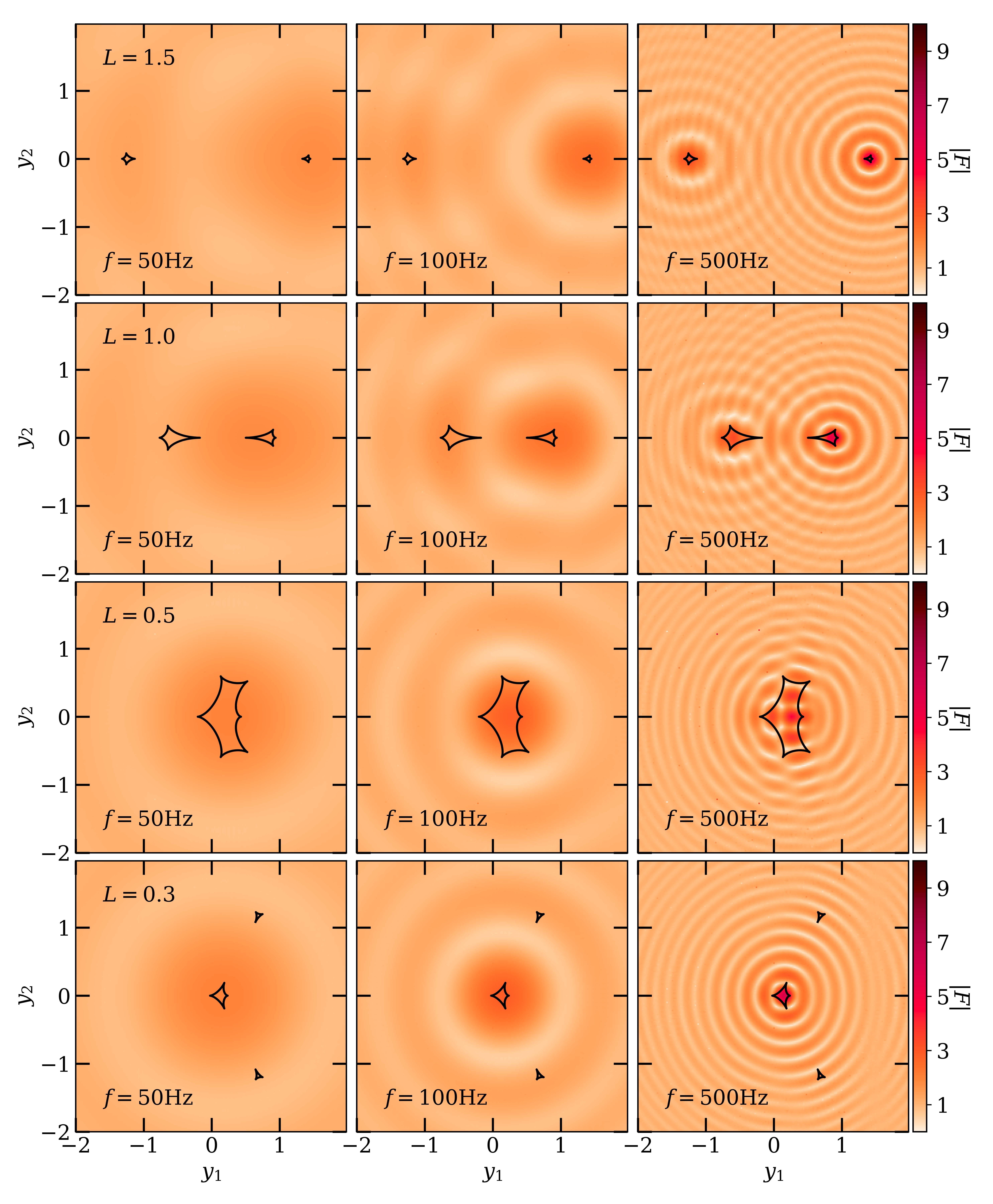}
    \caption{Amplification factor, $F(f)$, maps for binary lens with~$(q_1, q_2) = (0.75, 0.25)$ and~$M_T=200~{\rm M_\odot}$. In each row, the left, middle, and right panels correspond to~$f=50,100,500$~Hz. From top to bottom, we decrease the distance~($L$) of both point masses from the centre. In each panel, black solid curves represent the caustic structure.}
    \label{fig:bm_interf_comb2}
\end{figure*}

The interference patterns for a binary lens with equal mass ratio components and different~$L$ values are shown in Figure~\ref{fig:bml_interf_comb1}. Example amplification factor curves for some randomly chosen source positions, marked by a solid stars, are shown in Figure~\ref{fig:bml_Ff}. Again, as we have seen in the last section, amplification factor curves can differ significantly compared to geometric optics approximation close to caustics, where a number of virtual images changes~(top row in Figure~\ref{fig:bml_Ff}). As we move away from caustics, both the amplitude and phase of the oscillation start to converge towards geometric optics results~(bottom row in Figure~\ref{fig:bml_Ff}). For comparison, in the bottom panel, we also show the point mass amplification factor with~$y=|\pmb{y}|$. The interference patterns for a binary lens with a mass ratio ~$(q_1, q_2) = (0.25, 0.75)$ are shown in Figure~\ref{fig:bm_interf_comb2}. For a binary lens, again, the dimensionless frequency~$(\nu = 4{\rm G}M_{Tz}f/{\rm c^3})$, can be used to determine other pairs of total mass and frequency~(assuming the same mass ratios) which would produce the above amplification maps.

\section{Summary}
\label{sec:summary}
In this work, we simulate and look at the patterns in the amplification maps for simple lens models, namely, the point mass lens, the Chang-Refsdal lens, and the binary lens, primarily focusing on GW signals detected with the LVK detector network. For the point mass lens, we have a point caustic at the centre of the source plane. The corresponding amplification factor calculated under geometric optics closely follows the full solution after the first peak~(i.e., amplification region; see~\citealt{2022JCAP...07..022B}). However, once we go beyond the point mass lens, either by introducing external shear or by adding another point mass lens~(i.e., binary lens), the corresponding caustics cover regions of different multiplicities in the source plane. As the source moves close to the caustics, the corresponding amplification factor shows significant differences compared to the amplification factor obtained under geometric optics approximation. Caustics also mark the region around which we have multiple virtual images with similar and high amplification, and can lead to significant de-amplification around~${\sim}100$~Hz for a lens mass of~${\sim}100{\rm M_\odot}-200{\rm M_\odot}$. If these significantly de-amplified regions in the amplification factor align with ringdown frequencies in the LVK band, we may end up with GW signals, which may be hard to interpret without the incorporation of lensing.

Another interesting aspect would be to search for unique features in the amplification factor profile for different lens models. Previous works have discussed the importance of unique features in breaking local degeneracies and distinguishing different lens models~\citep[e.g.,][]{2024PhRvD.110j3024M}. For example, a lens model with a finite core will lead to the formation of a central image, which introduces secondary modulations to the amplification factor~\citep[e.g.,][]{2023PhRvD.108d3527T}. Although less likely for GW signals with typical signal-to-noise values, such features may allow us to infer the presence of lensing and the lens properties without doing detailed parameter inferences.

\section*{Acknowledgements}
The author thanks the anonymous referee for useful comments. The author acknowledges the support of grant 2020750 from the United States-Israel Binational Science Foundation (BSF), grant 2109066 from the United States National Science Foundation (NSF), and the Ministry of Science \& Technology, Israel. This research has made use of NASA’s Astrophysics Data System Bibliographic Services.

\bibliographystyle{apj}
\bibliography{Reference}

\end{document}